\documentclass[nonacm,sigplan]{acmart}

\usepackage[]{hyperref}
\usepackage{enumitem}

\begin{document}

\title{GigaAPI for GPU Parallelization}

\author{Mihir Suvarna}
\email{msuvarna@cs.utexas.edu}
\affiliation{%
  \institution{The University of Texas at Austin}
  \country{}
}

\author{Omeed Tehrani}
\email{omeed@cs.utexas.edu}
\affiliation{%
  \institution{The University of Texas at Austin}
  \country{}
}

\begin{abstract}

GigaAPI is a user-space API that simplifies multi-GPU programming, bridging the gap between the capabilities of parallel GPU systems and the ability of developers to harness their full potential. The API offers a comprehensive set of functionalities, including fundamental GPU operations, image processing, and complex GPU tasks, abstracting away the intricacies of low-level CUDA and C++ programming. GigaAPI's modular design aims to inspire future NVIDIA researchers to create a generalized, dynamic, extensible, and cross-GPU architecture-compatible API. Through experiments and simulations, we demonstrate the general efficiency gains achieved by leveraging GigaAPI's simplified multi-GPU programming model and showcase our learning experience through setup and other aspects, as we were interested in learning complex CUDA programming and parallelism. We hope that this contributes to the democratization of parallel GPU computing, enabling researchers and practitioners to unlock new possibilities across diverse domains.

\end{abstract}

\maketitle 
\pagestyle{plain}

\section{Introduction}
GPU's have existed since the cusp of the 21st century, dating back all the way to 1999 when NVIDIA first introduced the GeForce 256. Years later, GPU's have transformed into accelerators that can be used for high-power computing, being able to efficiently calculate complex algorithms and large volumes of data in parallel at higher speeds than CPU's. 

As the landscape for high-performance computing has been developing, there has been a shift from single to multi-GPU setups, where GPU's have begun to be used in parallel for even more efficient computing. This added parallelism has unlocked a plethora of problem-solving applications, with the ability to efficiently divide up tasks, compute on separate GPU's, and then combine results together. However, this increase in multi-GPU setups has introduced large levels of complexity, adding overhead when programming on such rigs and struggling to efficiently keep resource management simple. 
\subsection{Challenges in Multi-GPU Programming}
While multiple GPU's running in parallel offer computational benefits, the added complexity can deter users from being able to interact with them efficiently. Full utilization of two GPU's running in parallel requires extensive hardware and software knowledge, being able to understand how to run code with low-level CUDA in parallel, and can stretch beyond the complexity of what code is being written in user-space. 

Complexities like these not only prolong development cycles, but restrict the adoption of such large-scale systems in both industry and academia. Ensuring that the workload is evenly balanced in a distributed system, or even performing efficient memory management to avoid bottlenecks, prevents widespread adoption. While CPU to GPU transfer and integration has been extensively studied, the landscape for parallelism of GPU's is still relatively fresh, with more work to be explored in this area. The lack of standardization and portability that exists within today's frameworks (different vendors, different architectures, underlying proprietary technology) has translated to not fast enough progress on parallel computing. 

Focusing on a subset, CUDA \cite{cuda} has made large strides in addressing efficient parallel computing on NVIDIA GPU's. There has been lots of good work surrounding this, with large-scale research teams at NVIDIA developing advanced frameworks, and we look to build on such works with GigaAPI.
\subsection{Background and Related Works}
Parallel GPU programming has become crucial for accelerating computationally intensive tasks across various domains, whether it be image processing, basic operations (compounded in other tasks), or even machine learning. Massively parallel architecture of GPU's has enabled significant speedups compared to traditional CPU-based computing \cite{owens2008gpu}. Nickolls and Dally published a study in 2010 \cite{nickolls2010gpu} and discussed the emergence of the GPU computing era and the potential of GPU's for accelerating parallel workloads, optimistic about the future of what GPUs can bring for computing.

However, the depth and complexity of low-level CUDA and C++ programming hinders developers and even users from fully utilizing the potential of multi-GPU systems, as pointed out by Gregg and Hazelwood \cite{gregg2011where}. Lack of abstraction has led to frameworks being developed, specifically trying to solve the abstraction of machine/low-level code to userspace. 

One very popular framework, CUDA, developed by NVIDIA, has become the de facto standard for GPU programming \cite{cuda}. CUDA is very beneficial due to its platform, the research team at NVIDIA behind it, and even how large of a programming base it has. In this paper, we use CUDA 12.0, with very advanced toolkits and in-built libraries. As such, it has been a key player in the context of parallel GPU programming.

Prior to GPU-GPU pipelines, works explored systems for having CPU's talk efficiently to GPU's. One such work from Jablin et al. \cite{jablin2011automatic} was a proposed system for automatic CPU-GPU communication management and optimization through their CPU-GPU Communication Manager (CGCM). Gelado et al. \cite{gelado2010asymmetric} presented an asymmetric distributed shared memory model (ADSM) in 2010 (a year before Jablin et al.) for parallel systems as well, which aimed to provide a unified programming model across CPUs and GPUs.

As we reach beyond the space of just CPU-GPU interactions, GigaAPI looks to explore the GPU-GPU computing space, and builds upon existing works. We aim to provide a user-space API that simplifies multi-GPU programming, specifically when running two GPU's in parallel. The modular design of GigaAPI is inspired by the work of Owens et al. \cite{owens2008gpu} through concepts of parallelism and throughput, attempting to design a usable framework for all developers. 

Our API offers a comprehensive set of functionalities, including fundamental GPU operations, image processing, and complex GPU tasks. Through experiments and simulations, we demonstrate the general efficiency gains achieved by leveraging GigaAPI's simplified multi-GPU programming model. By abstracting away the complexities of low-level programming, GigaAPI enables developers to focus on high-level functionality and algorithm development, ultimately facilitating the adoption of parallel GPU computing in various domains.

\subsection{Simplified Multiple GPU Abstraction: GigaAPI}
As aforementioned, the gap that exists between the capabilities of multi-GPU systems and the ability for developers to harness these systems is large. A much more intuitive programming interface could offset challenges faced when using such systems, abstracting away the complexity and making it much more feasible to harness the full computing power of such resources in user-space. We propose a novel user-space API, called GigaAPI, which treats two GPU's as a "giga-GPU", simplifying the backend significantly.

Our abstraction encapsulates the lower level thread coordination, device synchronization, and memory management into a framework that essentially anyone can use. The emphasis on such a simple framework is twofold: it is easily extensible and open-sourced, while it is simple enough to digest for any machine running two GPU's in parallel. When we first set out developing this, we had one high-level idea in mind, which was to increase productivity and have developers focus on the algorithmic aspects of a project rather than the parallel intricacies. 

An important piece to state is that our project is merely a proof-of-concept of an API in userspace, with a demonstration of simple and complex tasks that we have developed. These range from basic operations that GPU's do on a frequent basis, to more complex manipulations on images, and then to even more complicated tasks. As some of the tasks we do are rudimentary, there have already been implementations by industry-grade researchers, including multiple CUDA research teams that have created parallelized code. 

In this paper, we explore how we setup and created such a framework, the low-level abstractions that enable operations to have low latency and high throughput in parallel, and how we benchmark such cases against existing SOTA approaches. 

\section{System Specifications and Setup}
Prior to designing GigaAPI, we needed to get access to two NVIDIA parallel GPU's, as well as a machine that is capable of running low-level CUDA code. Our initial thoughts was to just use our existing personal machines, and purchase an additional GPU to add in tandem, but this turned out to be much more complex than we initially anticipated. First, our hardware is all SFF (small form factor): that is, the motherboard, unfortunately, is mITX and has only one PCIe slot. Thus, we sought to find another such usable machine.

After speaking with the UTCS help department and Professor Rossbach, we were able to get access to \verb|pedagogical-7|, which suited our use case perfectly. The CPU included on this machine was a Intel(R) Xeon(R) Gold 6226R CPU, which is a Hexadeca-core processor (16 cores) with a speed of 2.90 GHz. We also had 99,270,656 kB of free memory, which is 100GB of usable memory. Past this, we have two GPU's running in parallel: a set of two NVIDIA Quadro RTX 6000's, which is a fantastic amount of compute. Running \verb|nvidia-smi| gave us information about the actual GPU's, what NVIDIA drivers they were running, and the available memory from each GPU. It should also be worth noting here that we used a tool called \verb|nvtop|, which is designed for showing readable GPU usage information \verb|top| style for NVIDIA cards, as we did not have \verb|sudo| access to these machines for more detailed usage stats. 

\subsection{CUDA Setup}
Setting up CUDA on the lab machine was our first roadblock, and the initial step that we had to perform. Since our code relies very heavily on the underlying CUDA operations, it's important that we have a suitable CUDA installation set up on these machines. Since the usual install location is somewhere in the root directory, we chose to install CUDA and the CUDA compiler (\verb|nvcc|) in our own home directories (e.g. \verb|/u/msuvarna|. We installed CUDA-12.0, which is the version compatible with our RTX 6000's. 

The setup was done as follows: first, getting the download through \verb|wget|, then making the binary executable, and then using the inbuilt GUI from the installer to have it install into a folder \verb|cuda-12.0| on our individual home directory. It should be worth noting here that since we did not have a shared directory, we had to install CUDA twice, but the versions and installs are the exact same (copied from one user to another). Multiple issues were run into when installing this, as we did not have \verb|sudo| access: we simply kept tweaking and modifying the install options until it only installed everything into our own home directory (requiring no permissions), and then it worked great. The last step involved exporting the new path to CUDA and appending it to our current path in our \verb|.zshrc| file.
\subsection{OpenCV Setup}
Past this, we had to install OpenCV as well, since our whole image API suite relies on this heavily for various operations. We performed the same operation of downloading OpenCV using \verb|wget| and created a separate directory (\verb|/opencv-install|). We then built the directory, created a sub-directory called \verb|opt|, then moved all the install files into this sub-directory, and finally ran \verb|make install|. This gave us all OpenCV headers and files in user-space, within our home directory: we did not find a suitable version on the lab machine, hence the need to install it this way.

\subsection{Remark about OpenCV and CUDA, cuFFT, cuBLAS}
As a lot of our benchmarks and baseline implementations later in the experiments section relied on OpenCV and existing CUDA libraries/scripts, we would like to give quick attribution to these resources. There is a great CUDA C++ Programming guide available at \href{https://docs.nvidia.com/cuda/cuda-c-programming-guide/index.html}{\textcolor{blue}{CUDA C++ Programming Guide}} and a fantastic cuBLAS user guide available at \href{https://docs.nvidia.com/cuda/pdf/CUBLAS\_Library.pdf}{\textcolor{blue}{cuBLAS Library Manual}}. Reading through lots of documentation available online helped us get quite familiar with how CUDA operates, the way it profiles GPU's, how it launches streams and kernels in parallel, etc. It was quite helpful and a great resource for debugging as well, as in the case we got stuck, we found error messages from the CUDA library and in-built helpers as well.

\section{API Design (high-level)}

GigaAPI, a unique name inspired by the NVIDIA GigaThread architecture and a playful nod to the internet meme "giga chad," is a user-space API implemented on a Linux system, leveraging the power of two NVIDIA Quadro RTX-6000 graphics cards. This section will explore the three main categories of functionality offered by GigaAPI, providing an in-depth look at the capabilities within each category and offering guidance on how users can effectively harness these features. Following this overview, we will delve into the lower-level details of our CUDA and C++ implementation, discussing the design decisions that have shaped the development of these functionalities. By the end of this discussion, users will have a comprehensive understanding of how to utilize GigaAPI to its fullest potential in their applications. Additionally, we hope that researchers will be inspired to build more general APIs with a wider range of applicability, enabling the power of GPUs to be harnessed in a technology-driven, GPU-centric world. 

\subsection{Fundamental GPU Operations (3)}

When we consider the term "fundamental" in the context of GPU parallelism, we are referring to a set of essential, foundational operations that form the basis for a wide range of applications across various domains. These fundamental operations are not only basic building blocks but also critical in terms of their importance and the significant performance benefits they offer when executed on multiple GPUs in parallel. After careful research, we identified AND chose \textbf{three} main categories of operations that meet these criteria: 

\begin{enumerate}[label=\textit{\arabic*.},ref=\textit{\arabic*}]
    \item \textit{Fast Fourier Transform (FFT)}
    \item \textit{Matrix Multiplication}
    \item \textit{Vector Operations (SIMD or Single Instruction, Multiple Data)}
\end{enumerate}

Let's break down the importance of these. Fast Fourier Transform (FFT) has numerous real-world applications across various domains. One of the most prominent areas is signal processing, where FFT plays a crucial role in audio compression, speech recognition and synthesis, audio filtering, and even spectrum analysis of audio signals. FFT has also been found to be used extensively in seismic and geophysics research, enabling efficient analysis and processing of complex geophysical data. Cryptography is another significant field that heavily relies on FFT, particularly in cryptanalysis, code breaking, and the development of provably secure hash functions like SWIFFT, which is entirely based on FFT concepts.

Moreover, during our research in graduate robotics labs at UT, we learned about a powerful FFT-based technique for evaluating signals detected by robotic system sensors. This technique effectively removes undesired signals or noise generated by vibrations from the surroundings, both when the robot is at rest and in motion, ensuring smooth and uninterrupted system functioning. By eliminating these disruptive elements, the FFT-based approach enhances the accuracy and reliability of the robotic system's balance, resulting in optimal performance and stability. Learning to parallelize these operations can have a vast impact on speed and performance, especially in real-world robotics systems, showcasing the need for efficient parallelization.

Matrix multiplication, a cornerstone of linear algebra, finds extensive applications across a wide spectrum of real-world domains. Its prevalence is particularly notable in computer graphics and animation, where it plays an important role in raytracing applications, camera projection, rotation, scaling, and other transformations (which we learned from Computer Graphics - Honors this semester). Machine learning, especially in the realm of neural network computation, heavily relies on matrix multiplication for efficient forward and backward propagation. Quantum computing also leverages this fundamental operation for qubit-related state transformations and operations. Public-key cryptography, a critical component of secure communication systems, is another such domain that utilizes matrix multiplication. 

In the field of robotics, matrix multiplication is instrumental in solving inverse kinematics problems, and for one of our final papers in another course, we implemented inverse kinematics and solved for a Jacobian matrix in 3D... which could potentially be parallelized by doing element wise or row/column wise parallelization (time permitting). Signal and image processing tasks, such as convolution and intersection with Fast Fourier Transform (FFT), also benefit from the power of matrix multiplication. 

Beyond these domains, matrix multiplication finds significant use cases in finance and economics, including option pricing, derivatives modeling, and portfolio optimization, where the speed and parallelism offered by matrix operations are essential. While this list is not exhaustive, it serves to highlight the pervasive nature of matrix multiplication and its indispensable role across a diverse range of applications that a user might want to develop in user-space. 

Vector operations, our third and final selection, have strong resemblance to matrix operations and share a lot of similarities. However, vector operations find more widespread use in scenarios where the dimensionality is constrained to one dimension. A key distinguishing factor is that vector operations do not necessitate complex manipulations, such as inversions, due to their 1D nature. They are simple and efficient, which makes them a great choice for audio signal processing where the data might consistent of a single dimensional time series, or collision detection in RayTracing, particle simulation, etc. While many implementations are still confined on CPUs, modern operations make work of the GPU and are highly optimized to run on parallel architectures, where batching and handling compute separately on two GPUs becomes trivial. 

Machine learning is another such domain in which vector operations are a key basis, used across a variety of algorithms: deep learning, include forward and backward propagation, rely on loads of vector operations and parallelizing these can make a model train more efficiently, even possibly saving compute time. Gradient descent optimizations are another such exmaple, with plethora of vector calculations included. Solving large systems of linear equations (in the realm of $Ax=b$) is another such example of how good vector parallelizations can have a strong impact. Hence, such a strong use case for such a simple operation necessitated a need to have parallel GPU's optimizing execution, improving performance and speed across the board.

\subsection{Image and Graphics Operations (3)}

In addition to the fundamental operations that form the building blocks for various applications, we also recognized the importance of providing a set of tools that enable users to develop more extensive and complex applications. These tools are designed to showcase the power and efficiency of parallelism in real-world scenarios. We chose three operations we believe to be fundamental in image processing pipelines:

\begin{enumerate}[label=\textit{\arabic*.},ref=\textit{\arabic*}]
    \item \textit{Upsampling} - increases the resolution of an image while not losing too much visual quality, such as going from 200x200 to 2000x2000 (for example).
    \item \textit{Sharpening} - enhances edges and details within an image to improve its visual clarity and make it seem more realistic with a boldness to it.
    \item \textit{Grayscaling} - converts a colored image to a grayscale representation. 
\end{enumerate}

Let's look at image upsampling for example, just to justify our decision. While splitting an image into pixels and parallelizing the upsampling operation is a straightforward task, the real challenge lies in scaling this process to handle large databases of images. The difference in processing time and resource utilization between upsampling a single image and handling hundreds or thousands of images is significant. By providing an API that enables efficient parallelism on a single image, we aim to demonstrate the potential impact on use cases where large-scale image processing is required. 

Consider a scenario where a non-technical UT Austin student teaches themselves to code and develops a mobile app called "CS380L Austin Gems" that showcases user-generated photos of hidden gems and popular landmarks in Austin, inspired by an past Advanced OS student. To ensure a consistent and high-quality user experience, they seek to implement upsampling as a feature. As the app gains popularity and the volume of image uploads increases, the student leverages parallel processing techniques to efficiently handle the workload, ensuring that every photo is quickly upsampled and displayed beautifully within the app's interface. Having an API to do this without getting knee deep into CUDA or C++ coding would be very beneficial to this student. 

The same principles can be applied to a wide range of image processing tasks, such as sharpening and grayscaling. Additionally, we did this out of interest to combine OpenCV with CUDA programming as well! 

\subsection{Complex GPU Applications (Theoretical)}
We explored some tasks within the realm of sorting, algorithms, and even machine learning, but found them to be \textit{much} harder than we initially anticipated. In lieu of this, we instead decided to focus instead on researching about these operations and how they could be parallelized. 

\begin{enumerate}[label=\textit{\arabic*.},ref=\textit{\arabic*}]
    \item \textit{Simulated Bitcoin Mining}
    \item \textit{Monte-Carlo Simulation}
    \item \textit{LLM's}
\end{enumerate}

One scenario that we were super interested in is the Bitcoin mining process. In this process, miners compete to find a hash that meets tha target difficulty set by the bitcoin network. The process involves generating a number of hashes by iterating through different nonce values and checking if the resulting hash meets the difficulty target. We attempted to do a simulated version of this by first creating a mining data file with very basic values that we could use to generate new hashes using the nonce values. 

We had a \texttt{parallelMining()} Kernel function that takes the blockData string and the target string as an input, generates a range of nonces and appends each nonce to the block data to create a set of the mining data. This is copied to the GPU memory using cudaMalloc and then the launch mining kernel runs on the GPU and processes each line of the mining data in parallel. For each line, it computes the hash using a simple hash function (not the actual hash used in bitcoin scenarios), and the result of each lines validation of the hash is stored in the results array on the GPU. After the execution of the kernel finishes, it iterates over the results array to see if any line had a valid hash in the network. 

We dealt with a lot of issues in this implementation, likely because of the hash and the way that we were distributing the work load. Also, there are not guarantees to find a target, and we likely should have set up some simulated block chain for the code to connect to, instead of simulating things out of a text file and making guesses on hashes and nonces. 

Monte-Carlo simulations rely on repeated sampling to obtain numerical results, and have shown strong use cases in finance, physics, and even in machine learning. They can help us model complex systems, maybe even concepts unknown to us.

With relating it to GigaAPI, Monte Carlo simulations can be parallelized quite efficiently (as hypothesized). Since the sampling that happens is random and independent, it is not hard to distribute the workload across multiple GPU's. Ideally, one GPU would generate its own set of samples, performing all computations, while the other GPU works in parallel to do the same, effectively (in theory) halving the time needed to compute. 

However, when attempting to implement this, we ran into loads of memory issues, where random samples would either oscillate and not converge, or just bad random number generators to begin with. Aggregating the results was also no easy feat: when combining what was batched from each GPU, the final collective sampling batch was not able to be translated into anything useful. We tried particle filtering, even some Markov chains, and even random sampling. Given more time, this is definitely something we wanted to explore much more in depth!

Our final approach was attempting to learn how to parallelize training an LLM. A very large issue with LLM's today is that for most consumers, models do not fit on the GPU. Either it is too big in memory, and doesn't fit on one GPU card, or consumer hardware is just not good enough to run such large models. Parallel GPU's offer an effective solution here: we can effectively split the model across two GPU's, distributing and balancing out the load.

Training such LLM's can be accelerated by using a parallel and distributed process. Essentially, our plan was to replicate the model by having each GPU process a subset of the training data, utilizing the parallel architecture we have. Then, computing the gradients and aggregating them helps update the model parameters, reducing the training time. 

With our other GPU, we can also utilize a testing process that evaluates the model performace: generating predictions or maybe even evaluating a held-out dataset would be a good option here, as we have the spare compute to dedicate. We even considered doing some sort of train/test loop, where we chunk the model at savepoints and test it on our other GPU, allowing for concurrent training and evaluation.

However, implementation is a completely different story. Either there are already amazing implementations that exist for this task (frameworks Alpa and Ray from NVIDIA) or the communication between the GPU handoff outweighs the benefit of having two separate GPU's. This is definitely a part of the future works section: one that will be attempted sometime soon!
\section{API Design (Low-Level)}

To fully understand how GigaAPI works, we will dig into some of the lower level details of this implementation. 

\subsection{Repository Folder and File Structure}

\begin{enumerate}
    \item \#\_functionality directory: Main development occurs in the parent/root directory, and then once complete, we branched out each functionality into its own respective folder with a \verb|Makefile| which can be executed as desired. 
    \\
    \item \verb|CudaKernels.c|: At a high level, this file contains CUDA kernels that perform various image processing and mathematical operations. Kernels include all 3 subgroups of tasks:
    \begin{itemize}
    \setlength{\itemindent}{+2em}
        \item Fundamental GPU Operations
        \begin{itemize}
            \setlength{\itemindent}{+2em}
            \item Fast Fourier Transform
            \item Matrix Multiplication
            \item Vector Operations
        \end{itemize}
        \item Image Operations
        \begin{itemize}
            \setlength{\itemindent}{+2em}
            \item Rgb2Grayscale
            \item Upsampling
            \item Sharpening
        \end{itemize}
        \item \textbf{Attempted} Hard Tasks
        \begin{itemize}
            \setlength{\itemindent}{+2em}
            \item Simulated Bitcoin Mining
            \item Monte-Carlo Simulation
            \item LLM training
        \end{itemize}
    \end{itemize}
    These kernels are defined at \verb|__global__|, which is used to declare the scope, and then is invoked using \verb|<< >>| syntax, specifying the grid and block dimensions for execution with respective launch kernel functions. See section \ref{sec:4.2} for more.
    \\
    \item \verb|GigaGPU.cpp| and \verb|GigaGPU.h|: These are the main API header and API declaration files. The header file describes the variables and functions (basic interface) of the GigaGPU. The GigaGPU implements the actual files that handle all the low-level and backend operations for what happens when an API call (some function) is invoked. It creates a class out of the GigaGPU object in an object-oriented fashion, then has functions that can be called in an API fashion. See section \ref{sec:API} for more info.
    \\
    \item \verb|Makefile|: This is an adaptable and baseline 
    \verb|Makefile| that we generated. It includes the \verb|nvcc| compile options, compiler options for \verb|gcc|, and the linker specifications for all the OpenCV and CUDA-related libraries that we used.
\end{enumerate}

\subsection{Functionality Design and Implementation}
\subsubsection{Cuda Kernels} \label{sec:4.2}These are essentially the low-level operations that are taking place on the GPU, including defining the kernel with thread operations, creating a method signature (callable from C++, keyword \verb|extern|) for launching said kernels, and defines blocks of threads (covering the whole input) with what these thread blocks are performing.

Kernels follow a very similar structure, as they are all distributed to compute across two GPU's (modularized as well, easy to call), specialized for certain tasks, and utilizing the usable GPU resources. We chose to go with a pretty standard size of blocks that are 16x16 in size for most of our kernels (defined to be a block of 256 threads total). Extending upon what we have learned in class, this evenly divides into 8 warps onto our GPU's, running in parallel for a total of 16 warps running while both GPUs are up. 

16x16 was a design decision that we made: we found that there was not much of an advantage to increasing the block of threads and the balance was that a 16x16 block was good enough to be a cover for most inputs (whether that was matrices, images, etc.). 8 warps achieves a good performance when running tasks in parallel, allowing us to utilize the GPU well (not under or over). While this block size maximizes performance, it also balances out compatibility amongst other GPU architectures: we reason that such a block size should be suitable to implement this API on other such architectures, limiting the tweaks needed to make it function as intended. 

\subsubsection{API Header and associated \text{.cpp} files} \label{sec:API}
As a quick preface, this is just about how we expect our API to be used. Looking at this from a very high level, we aim to have users include \verb|GigaAPI.h| as a header file at the beginning of a fresh \verb|.cpp| file, with which all related files and functions come with it.

As mentioned before, we took an object-oriented approach to this: the user gets access to a \verb|GigaGPU| object, with all it's related functionality and tools. As a second note, there is also internal documentation presented with the actual API bundle itself, explaining function signature, constructors and destructors, etc. time permitting, we hoped to get all of this up and running on an EC2 instance, in which users could just use \verb|wget| to access and download all necessary libraries and headers in realtime, but unfortunately, we did not pay our AWS bill, and our container was revoked. We even attempted to put our API on Firebase, but ran into some functionality problems, and scope this to a future work. 

We also made our directory structure reflect as much as possible of the original NVIDIA CUDA Samples repository, in which there are folders that explain all the different functionality and associated example \verb|.cpp| files that can be run. To this extent, we created folders in our own repository (titled \verb|0_images| and \verb|1_basic|), in which each folder has three more subfolders of each and every example. There is an associated \verb|Makefile| with all of them, making it really easy to compile and run examples, as well as learn how to modularize and use the code outside of the repository without much difficulty.
\subsubsection{Upsampling}

This upsampling implementation promps the user to choose between upsampling a single image or upscaling a database of images. Based on the users choice, we call upon two possible functions, \texttt{processSingleImage()} or \texttt{processPhotoDatabase()}. The process single image function prompts the user to enter the path to a singular image and their desired scaling factor. Next, using those inputs, it calls \texttt{runSingleImageUpsampling()}, which takes the image path, reads the image using the OpenCV library, converts it to a vector of unsigned chars, and them passes it to a GigaGPU class function called \texttt{upsampleImage()}.

The \texttt{upsampleImage} function is the main entry point for upsampling the image. It takes the input image data, output image data, image dimensions, and the scale factor as parameters. It first calculates the size of the image data for each GPU, assuming that the image will be split into two equal parts so that each GPU can process half of the image. Memory is allocated on both GPUs using \texttt{cudaMalloc()} to store both the input and output image data. The input image data is then split into two halves, with each half being copied to a different GPU using \texttt{cudaMemcpy()}. 

The code calls the \texttt{launchUpsampleKernel()} function twice, once for each GPU. This function, in sequence, then launches the \texttt{upsampleKernel()} on the correct GPU, which performs the actual upsampling using a technique known as nearest neighbor interpolation. 

In the "nearest neighbor interpolation" approach used in this code, each thread in the CUDA grid is assigned a unique \texttt{(x, y)} coordinate based on its thread index and block index. The threads check if their coordinates fall within the dimensions of the upscaled image. If they are within the valid range, the source pixel coordinates in the input image are calculated by dividing the upscaled coordinates by the scale factor.

For each color channel (assumed to be RGB), the thread copies the color value from the source pixel in the input image to the corresponding pixel in the output image. This process is repeated for all threads in the CUDA grid to upsample the entire image.

A good way of putting it is that we use a "flavor" of nearest neighbor interpolation, but we don't actually do any interpolation from a mathematical sense. We effectively copy the color values from the nearest corresponding pixel in the input image to the upscaled image, without performing any interpolation or averaging of neighboring pixels. Although bilinear interpolation is a more popular and common technique for image upsampling, it was not implemented due to time constraints, and we highly recommend this.

Finally, after the kernel launch is complete on both GPUs, the \texttt{cudaStreamSynchronize()} function is called... which is essential to the concurrency here. We do an asynchronous memory copy for both devices as well. We additionally synchronize the devices. The output data from both GPUs is concatenated to form the complete upscaled image, and memory is freed, respectively. Below is the result of a sample run with 15x scale factor. \texttt{Image.png} is simply an image of Mihir and I at his birthday party. As shown in the photo, the image effectively upsamples from 294x406 to 2966x4096, with not a lot of loss in quality at all. We also tested this on three different versions that we wrote to make sure it worked for benchmarking, an OpenCV (CPU) version, a single GPU version and a dual-GPU version.  \\

\includegraphics[width=1\linewidth]{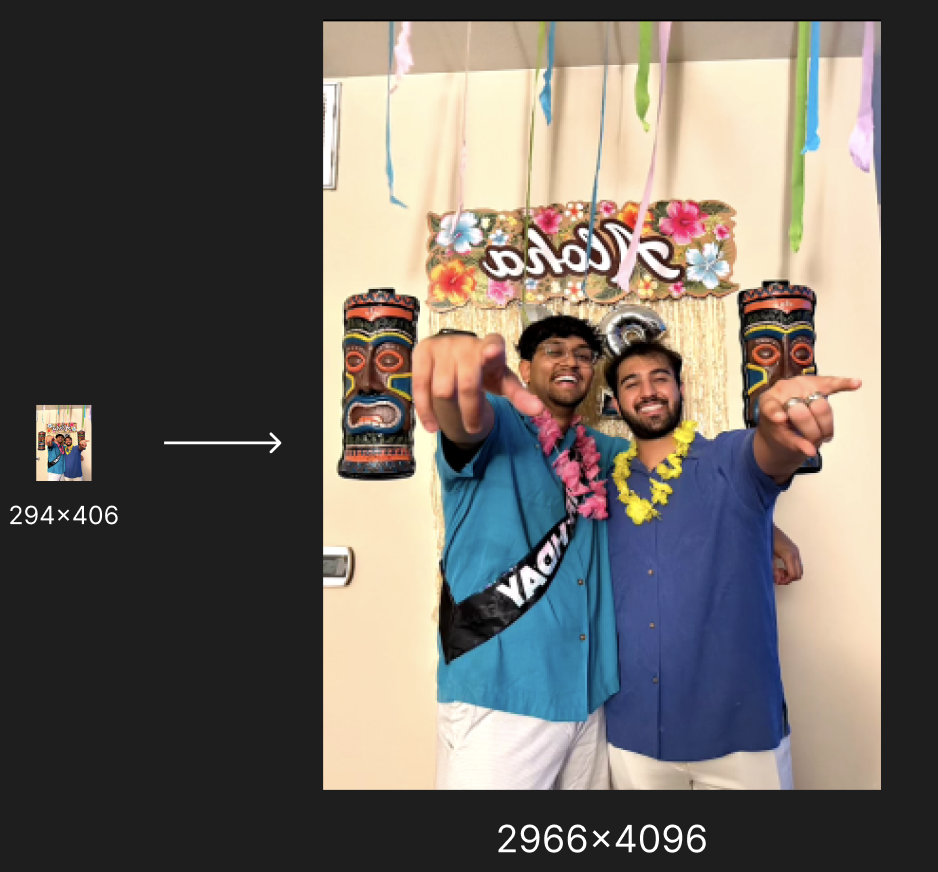}

\subsubsection{Sharpening}

For the sake of not re-explaining, the overall structure and flow of the \texttt{sharpenImage()} function remain the same as the \texttt{upsampleImage()} function. It first calculates the size of the image data for each GPU, does the allocation on the GPU, copies the data, launches the kernel, synchronizes operations, and copies the data back. The real fruit of the design is in the kernel function. We use the Laplacian filtering technique!

In the field of image sharpening, there exist various techniques beyond the Laplacian filter, each with its own strengths and characteristics. These methods include Prewitt and Scharr operators, high-pass filtering, and unsharp masking, among others. However, for the purpose our class project, the Laplacian filter was chosen due to several reasons. Firstly, the Laplacian filter benefits from a large amount of online resources. Secondly, the Laplacian filter stands out for its relative simplicity in implementation, which is advantageous when working with parallel processing frameworks like CUDA. Finally, and perhaps most importantly, the Laplacian filter excels at highlighting the fine details and edges within an image, providing a clear and immediate visual indication of its effectiveness. This allowed for us to quickly validate the correctness of the sharpening process and ensure that the desired outcome is being achieved. Consequently, the Laplacian filter emerged as the most suitable choice for this particular study, balancing accessibility, ease of implementation, and visual clarity in the sharpened results.

To begin, we define a constant array of floats that represents the filter kernel used for sharpening the image. The values in the filter are carefully selected to enhance the high-frequency components of the image, resulting in a sharpening effect. There are two commonly used discrete approximations to the Laplacian filter, and in this case, we opted for the second one, which consists of -1's surrounding an 8 in the center.

Inside the kernel, each thread calculates its unique thread ID based on the block and thread indices. This thread ID determines the specific pixel coordinates that the thread is responsible for processing. The kernel then checks to ensure that the calculated pixel coordinates fall within the valid dimensions of the image.

Next, we calculate the index of the current pixel in the image data array and iterate over each color channel of that pixel. For each color channel, we compute the weighted sum of the neighboring pixels. This is accomplished by utilizing two nested loops that cover a 3x3 neighborhood centered around the current pixel.

As long as the neighboring pixel falls within the image boundaries, the kernel calculates the index of the neighboring pixel in the input image data array and the corresponding index in the Laplacian filter. It retrieves the pixel value from the input image data array and the corresponding filter value from the Laplacian filter. The kernel then multiplies the pixel value by the filter value and adds the result to the sum variable.

We will delve into this further in the limitations section, but it's important to note that our current strategy does not fully exploit the parallelism available on the GPU since we iterate over the color channels sequentially. During the implementation phase, we encountered several bugs, particularly one related to neighbor indexing using variables i and j, which initially resulted in a smoother image instead of a sharpened one. We were surprised by the level of precision required in the calculations to achieve accurate results. While we believe that further exploiting parallelism could have showcased the deeper advantages of an API like this, especially if developed by a CUDA programming expert, our debugging experience highlighted the complexities and trade offs involved in such implementations.
\subsubsection{Grayscaling}
Again, we have a very similar structure that happens here: we have a \verb|convertToGrayscale()| operation available from our GigaGPU object, that can be easily invoked and called. As it stands, we keep an image inside the \verb|grayscale| folder, and our test image here was just an image of a Tesla Cybertruck charging a Tesla Model 3. That can be seen in our repo and has the filename \verb|cybertruck.jpeg|. It will then output an image that appends to the current filename with \verb|_grayscale.jpeg|, which will be the initial RGB image converted to full grayscale. 

The main entry point is \verb|convertToGrayscale| and it takes in the input image data, output image data (matrix for holding this), and the image width and height. First, we calculate the size of the input image that will be transferred over to each GPU: as you can imagine, some rounding does need to happen in the event of uneven pixels, and also all 3 channels of RGB need to be considered. Then, we allocate memory to each GPU using \verb|cudaMalloc()| to store the input and output image data. Finally, we split the image as before (based on the height) and then put each half on a different GPU. 

As before, we have a standard workflow: copy over the needed memory onto the respective GPU, launch the kernel on both devices, then synchronize and copy the results back. Inside the kernel, we calculate the unique ID for the block and thread indices (same as before) and it gives us the specific pixel location that will be converted from RGB to grayscale. We perform that operation using a simple calculation, in which we first make sure that the thread is within the bounds (as the blocks might exceed the image dimensions). Then, we individually extract the RGB values, using the coefficients (0.299, 0.587, and 0.114) to weight each color's contribution to the perceived brightness of the image. Finally, that calculated value is stored in the output image at the correct position, ensuring that every pixel is processed in the image.

When implementing this, we also ran into several issues, mostly along the lines of having not enough memory be allocated by the CUDA kernels. We had a very interesting bugs where certain image sizes would ultimately crash the kernel: images sampled at 8K UHD (as opposed to the standard 1080p HD we were working with) would not be full grayscale, which we deduced to being how we were dividing up the half height of the image. Another bug we ran into was how sometimes, channels presented an issue: if we did not use OpenCV to properly read in the image and assign it as RGB, the channels would sometimes be improper, messing with our grayscale images and giving it some sort of hue from time to time. Here's what our input test image was:
\begin{center}
    \includegraphics[width=.4\textwidth]{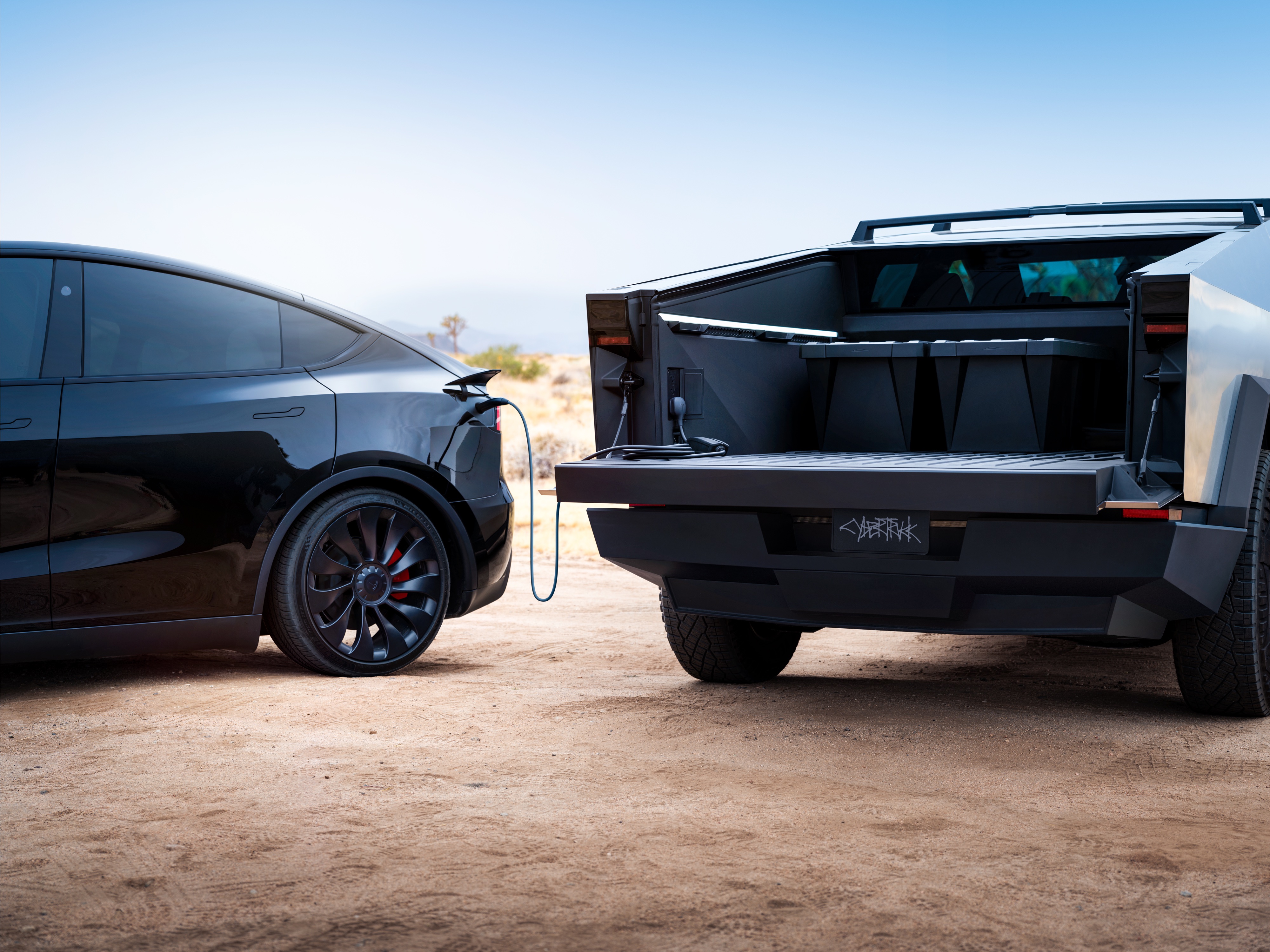}
\end{center}
And here's what our grayscale image looked like:
\begin{center}
    \includegraphics[width=.4\textwidth]{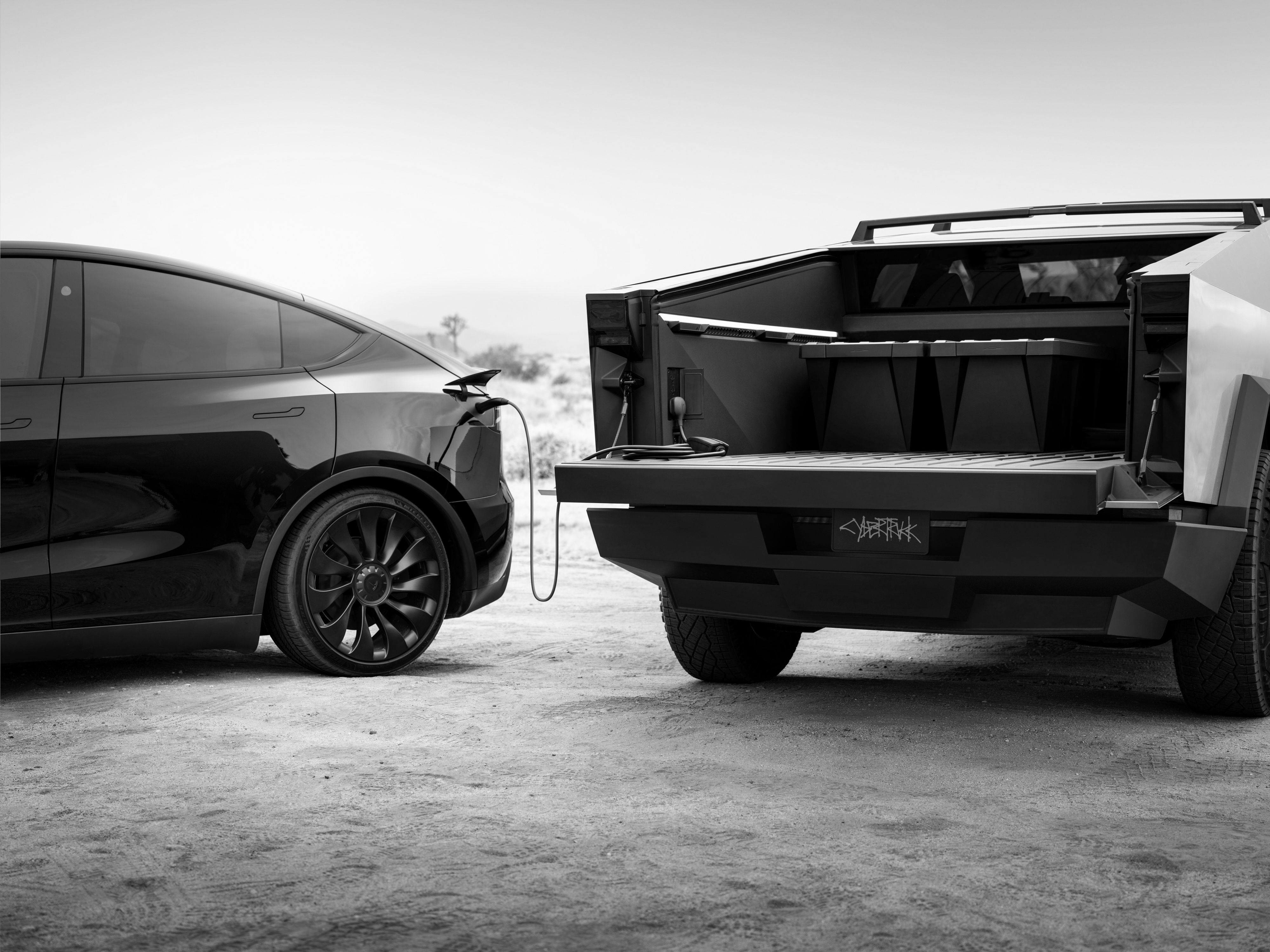}
\end{center}
We did run other image grayscale tests as well, but we thought this one was pretty interesting. It maintained the brightness of the sky quite well in grayscale, while also keeping relative colors of the actual vehicles themselves. There are still future improvements that can be done on this code itself, as it is very simple and just batches the image by half pixels on one GPU and half the pixels on another GPU: this is fine for relatively small images, but can blow up in complexity when dealing with either a batch of large images or some incredibly large RAW files themselves. We ran it with a RAW image file from a Nikon D7200 and was amazed at how much better it was parallelized, but also how long it still took due to the image size. It is a very simple operation that is being conducted, but it is pixelwise, hence the slowdowns we were seeing with much larger images.
\subsubsection{FFT}

There are a couple of pieces associated with the section of this project. For one, we provide a Python script to allow the user to simulate the creation of the raw data for a sinusoidal signal, which is a very common and fundamental signal in the context of FFTs.

Using some basic parameters (frequency, sample rate and duration), we create an input signal. The important value here is a sample rate of 1024, which can be altered. Here is a generated plot of the signal: \\

\includegraphics[width=.85\linewidth]{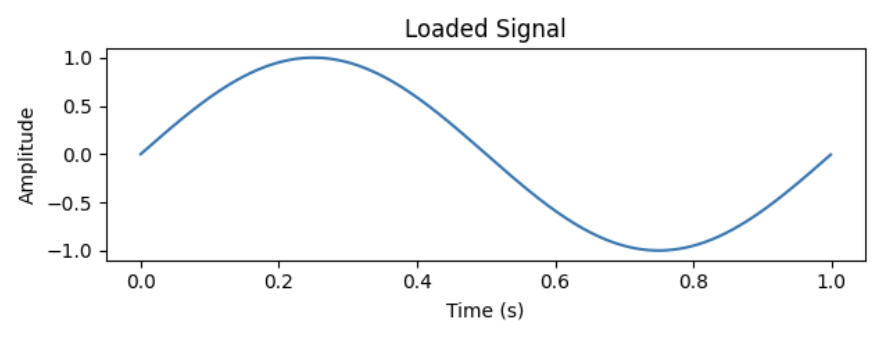}

Obviously this is purely for testing purposes. In a real world scenario, you'd want the user space API to be able to process a variety of different signals, and potentially intelligently be able to distinguish between them.

In the main method, we call upon a \texttt{prepareData()} function in our GigaAPI class, which prepares the input data for our \texttt{performFFT()} function. 

To perform a single GPU FFT computation, we create a cuFFT plan using the \texttt{cufftPlan2d()} function, specifying the dimensions and \texttt{CUFFT\_R2C} for real-to-complex FFT. We set up a stream for the plan to allow asynchronous execution and execute the plan to perform the computation on the GPU. In the API function call, we allocate the necessary device memory using \texttt{cudaMalloc()}, copy the input data from the host to the device, call the kernel function, and copy the output data back to the host. This approach focuses on utilizing a single GPU, as piecing together all the NVIDIA interfaces can be challenging when working with multiple GPUs.

Next, we begin to write a parallelized version of the FFT algorithm. First, we divide the input data into chunks based on the number of GPUs, which is two in this system. We realized that we should have allowed for parallelization across multiple GPUs beyond three to make this API as generalized as possible. We experimented with this idea here but not with the other functions.

We allocate memory on each GPU for the input and output data, copy the data from the host to the GPU, and then create separate streams for concurrent execution of the kernels. We call the \texttt{performFFTKernel()} function, which uses the same interface logic as described in the previous section. It's important to note that \texttt{cufftPlan2d()} is a single-GPU operation, so it's our responsibility on the API side to parallelize the computation.

After the computation is finished, we copy the output data back from the GPU to the host and free the respective memory. This chunk division works well in the case of FFT because the FFT algorithm divides the input signal into frequency components that can be computed independently of all other components. This allows for parallel computation of those frequency components without any dependencies between them. We really appreciated the NVIDIA interface code here, it shows how useful and easily to implement an API like this could be for a range of GPU tasks!

Finally, we open a file to write the output, and write each elements real and imaginary part to the file. This is because in FFT, the output consists of complex numbers which have a real and imaginary component (aka. \texttt{a + bi}). 

Here is an example of the output after the FFT completes:

\includegraphics[width=1\linewidth]{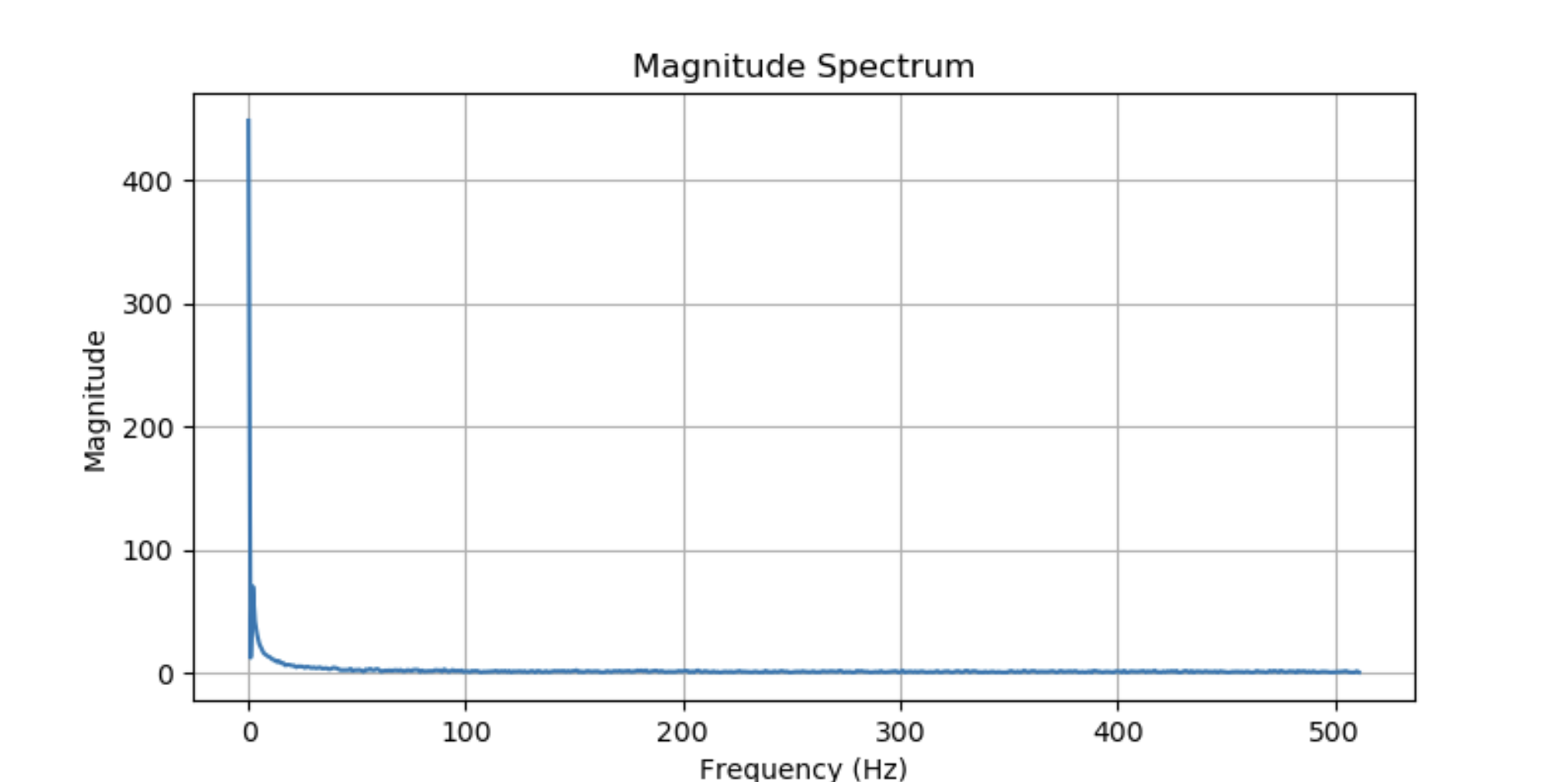}

\subsubsection{Matrix Multiplication} 
This structure was quite interesting to both code and make parallelized as well. Matrix multiplications, as aforementioned, are a very common task that are run across a variety of GPUs, and are often used as a low-level operation in much more complex tasks. So, we made our code quite modular and open to re-implementation, as in the future, if need be, the implementation can easily be changed to fit the use case. 

We start with our entry \verb|performMatrixMultiplication()|, in which this first does some elementary calculations to understand the matrix we are working with. It should be worth noting that we consider 3 matrices: A, B, and C. This is all configurable in the \verb|main| file, where one is using our API commands, and we initialize 3 matrices of any size to be any value.

Then, our program hands off control to the function, which first gets the sizes of all relevant matrices. Once the sizes are acquired, we effectively half them to parallelize the operations: half the multiplications are carried out on one GPU, while the other half are on the other GPU. First, memory is allocated on the device, then we copy over half the matrices, and finally start a stream individually on each device to handle computation. Once everything is copied over, the next step is to launch the actual CUDA kernel itself.

Our CUDA kernel is quite simple here: nothing far too complex, as we imagined our matrix multiplication to be quite applicable across a variety of tasks in future use cases. So, what we essentially do is that we have a kernel that computes a dot product: each element in C[i,j] is calculated by taking a dot product of the $i$-th row of the matrix A and the $j$-th column of matrix B. As before with other operations, row and column indices are calculated using thread indices, and that is how elements are computed independently on each GPU. 

If a thread's row and column is within the bounds of the matrix, then the dot product is computed, with the total sum being reported and assigned at the end of the loop across the row and column. These kernels are launched in the same fashion as before, where CUDA streams are launched to enable parallel execution of the code that is being run on each subsequent GPU. Same idea again: allocate memory, launch CUDA kernel, launch CUDA streams, synchronize and copy back the results, and finally free any used memory.

One very interesting thing to note here was during implementation, we had some issues with dirty memory. Upon finishing the execution of the matrix multiplication, we sometimes did not free the memory, and noticed initially that our values would often overlap from previous computation (which was a very weird bug, but expected). So, we had to do an in-depth memory check to truly understand what was going wrong, and after freeing all the memory, we noticed it was running as expected. Our test file is pretty simple: a function is even included to both initialize and print matrices, then runs everything through the high-level method (\verb|performMatrixMultiplication|) and passed to all the CUDA kernels. It works well for even large-sized GPU's: our tests went all the way up to a matrix of size 50,000 x 50,0000 with parallelization showing a noticeable speed-up in computation.
\subsubsection{Vector Operations}
Vector operations, as aforementioned, lie at the heart of every complex GPU operation as a simple function. It could be a variety of things that the user is doing: vector operations are almost always going to be running in the backend. Due to the complex nature of the variety of ways that vectors can be structured and stored, we chose to specifically focus on 1D vectors of varying size in GigaAPI, with a large emphasis on the L2 norm and the dot product. 

Data parallelism was a huge push for why we chose to focus on vector operations. When we were building this API functionality, we wanted to do a deeper dive into how it could be used from a user perspective: we thought a lot about large vector operations and how useful it would be to parallelize splitting data chunks across GPU's, specifically in a linear 50/50 index chunk. This leads directly into more thoroughput, as there is less chance of the GPU blocking trying to get through the whole array, and a bigger chance of more utilization (as seen in our testing using \verb|nvtop| to visualize the outputs). Data transfer times was a bit of a limitation here as there is some I/O needed to transfer from disk to hardware, and we admit that cuBLAS has already done a quite good implementation, but we wanted to get as close to that as possible and prove that our userspace parallelism would work on vector operations.

Our entry point into this function is twofold: for dot products, the API is called through the function \verb|computeDotProduct()|, while for L2 norm, we use the entry function \verb|computeL2Norm()|. Both operations are quite similar in nature: they first calculate the size of the data needed to be copied over onto the GPU (halving, with remainder going on one if not even split), begin streams on both GPU's, and then launch the respective kernels. 

We have very interesting kernel functions going on here. For the dot product kernel, we essentially calculate a running sum to accumulate the partial dot product, and we have a shared cache that is shared amongst all threads in the block (so all 256 threads, in 16x16 block). We process all threads by using a load-balancing technique, ensuring to cover the full input of the half of the array each GPU is tasked with computing for. Synchronization is used to ensure that threads finish writing to the cache before proceeding, preventing dirty writes and inconsistent memory accesses. Finally, we use a reduction to sum the partial sums (adding threads through halves of any active threads), and then report the final dot product. 

In that vein, the L2 Norm is almost entirely the same: it just deals with one vector, however, and then square roots the final result. Since the final part is just a total square root and is an operation performed only once, that is handled in the \verb|GigaGPU.cpp| file (after the kernels have finished and the streams are synchronized). Such an efficient approach to both ensures that we should get performance akin to state-of-the-art existing approaches: more about this will be explained in the results section.

\section{Design Limitations and Issues: A Discussion}

There are a large amount of limitations that we had for this project. We describe them in detail to ensure that anyone in the future who is inspired by this work can make a robust, general purpose user-space GigaAPI.

\textbf{Bottlenecks for Setup:} One of the significant challenges we encountered during the project, which was already under a tight timeline, was setting up OpenCV and CUDA on our accounts for the pedagogical-7 machines. This process proved to be time-consuming and added an extra layer of complexity to our already constrained schedule. Moreover, both of us faced disk space limitations on our accounts, further compounding the issues we had to navigate. To resolve this, we had to clear out files from our home directories and empty the cache, in order to free up the necessary storage space. 

\textbf{Fixed Number of GPUs:} Our GigaAPI code currently makes the assumption that the system has precisely two GPUs available, lacking the dynamic adaptability to accommodate a varying number of GPUs on a given system. Implementing this flexibility would have been a straightforward process, as demonstrated when we later extended the FFT functionality to leverage the NVIDIA interface for parallelization. During this exploration, we experimented with the concept of dynamic GPU adaptation to showcase its feasibility. However, due to time constraints, we did not fully integrate this feature into our repository. The implementation would have involved a simple approach of performing memory operations within loops and providing users with the ability to specify the number of GPUs present on their system. We would've have to adapt this to every function we wrote. This information could be passed either through a dedicated function call or as an input parameter to the relevant function.

\textbf{Inconsistent/Limited Error Handling:} Our error handling approach throughout the codebase lacks comprehensiveness, particularly in terms of memory checking. Given the extensive data transfer involved in GPU programming, a more thorough error handling strategy would have been beneficial to ensure robustness and reliability. 

\textbf{No Batch Processing for Images:} In our image processing functions, such as upsampling, the code handles only one image at a time. As outlined in the high-level API design, implementing functionality to process a database of images would be more beneficial. We attempted this but encountered numerous bugs when splitting batches and reusing kernel code for each image. Avoiding user iteration and utilizing GPU batching is challenging but could be a valuable future direction for GigaAPI. 


\section{Benchmarking and Evaluation}

\subsection{General Benchmark \#1: Is Parallelism Working?}

One of our key benchmarks in this project was to get the functionalities that we built out on our GigaAPI to actually work with the 
parallelism code, especially since we mainly did this project to try and learn CUDA programming, parallel programming, and how it all intertwines with a 
project and building out something like our user-space API in C++. First, lets take a look at a result from NVTOP, a classic command line tool that provides real-time information about the GPUs. To capture the event, we simply screen recorded the terminal and ran a very large image upscale:  \\

\includegraphics[width=.9\linewidth]{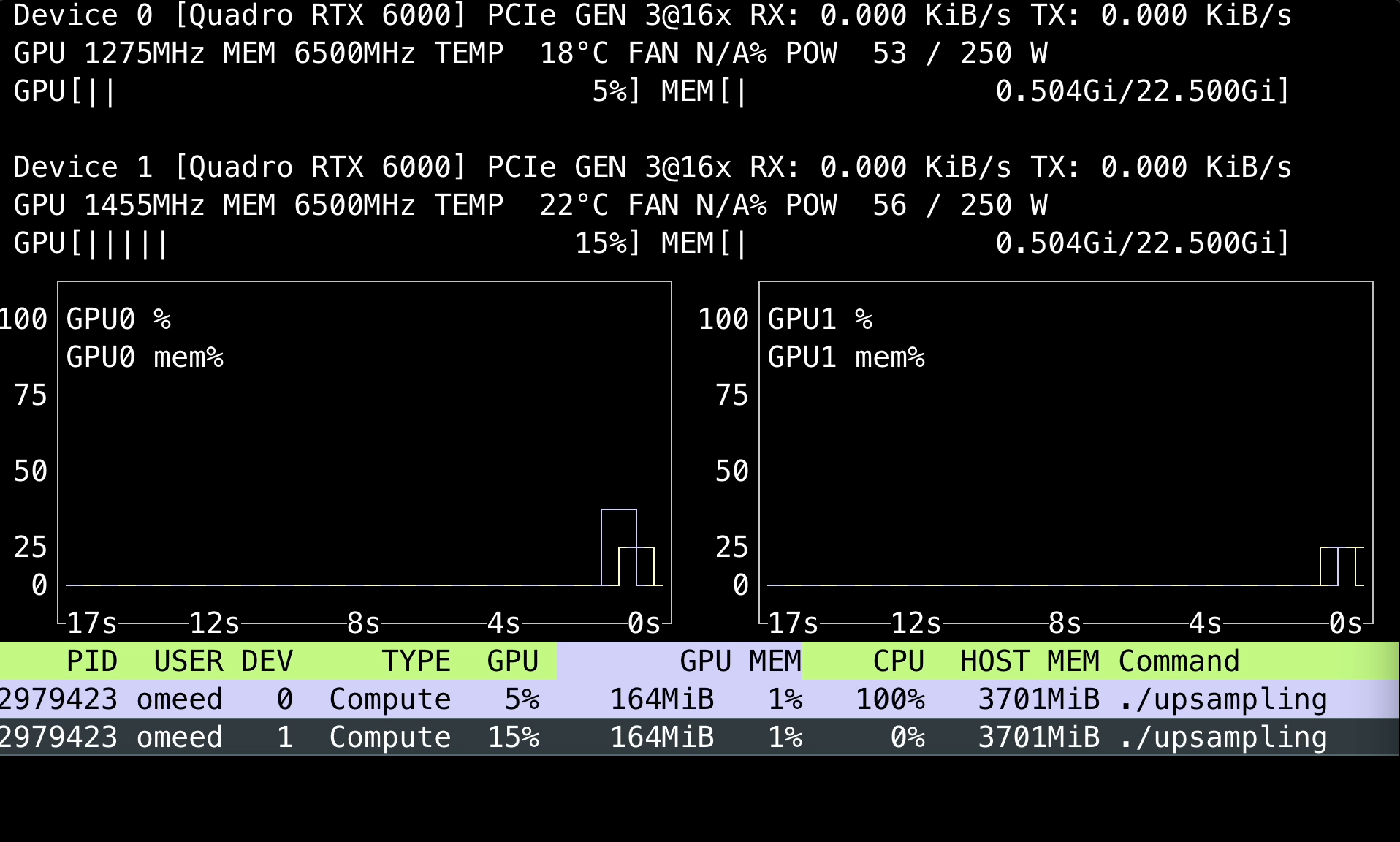}

The image above shows clear evidence of concurrent processing across multiple GPUs. The information presented reveals that process ID (PID) 2979423 is actively running on both Device 0 and Device 1, demonstrating effective workload distribution between the two GPUs. Furthermore, the compute row of the image displays the GPU utilization percentages, with Device 0 operating at 5 percent and Device 1 at 15 percent. These metrics confirm that the GPUs are actively engaged in parallel processing.

\subsection{API Benchmark \#1: FFT}
FFT is an interesting application of GPU parallelism. As previously noted, we know that cuFFT libraries exist, with the ability to perform FFT on signals at a very low latency, incredibly fast. For this benchmark, we were interested in testing out four commonly used signals: sine wave, sawtooth, square, and chirp. We used a Python helper script to generate the signals. Here is what those signals look like:

\begin{figure}[!ht]
    \centering
    \includegraphics[width=0.8\linewidth]{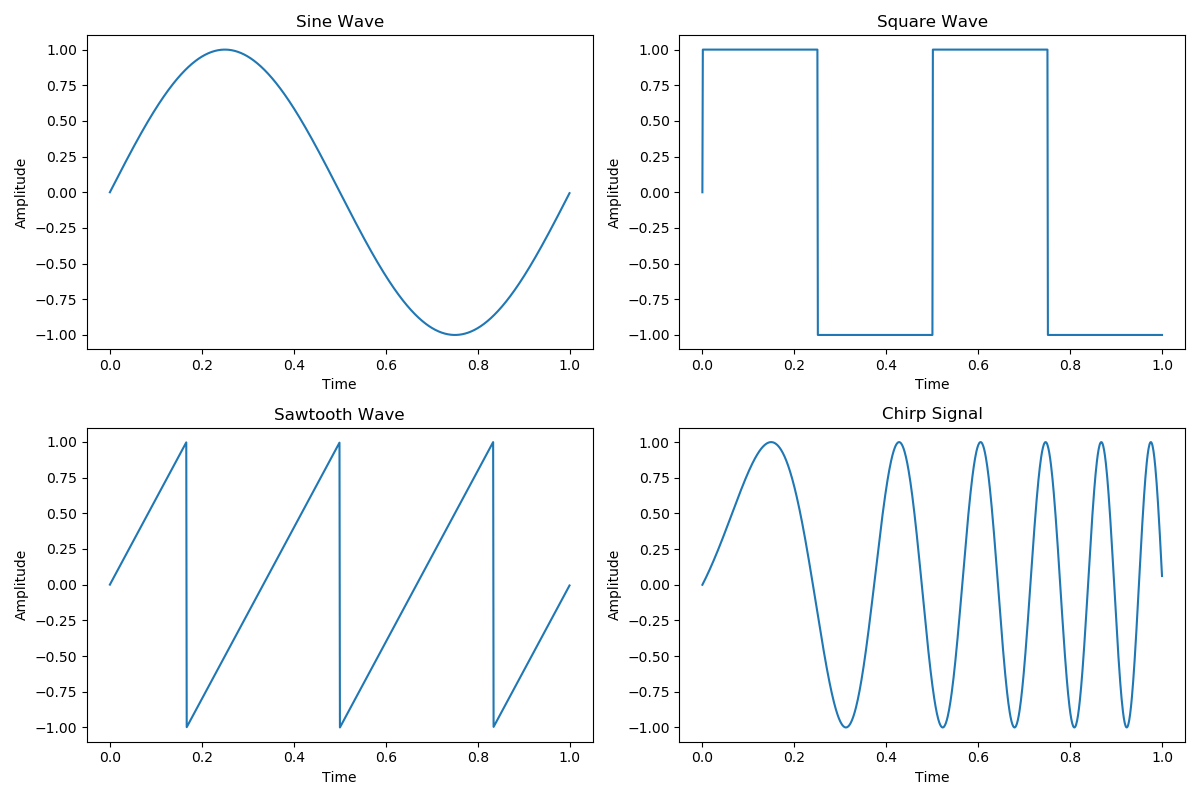}
    \caption{Pre-FFT Original Signals}
    \label{fig::fftwaves}
\end{figure}

After that, we wrote some baseline cuFFT single GPU code as our benchmark. We dug a little deeper into this library as well: cuFFT is quite well implemented, and we expected it to outperform our approach. Our benchmark results were as follows:

\begin{figure}[!ht]
    \centering
    \includegraphics[width=0.8\linewidth]{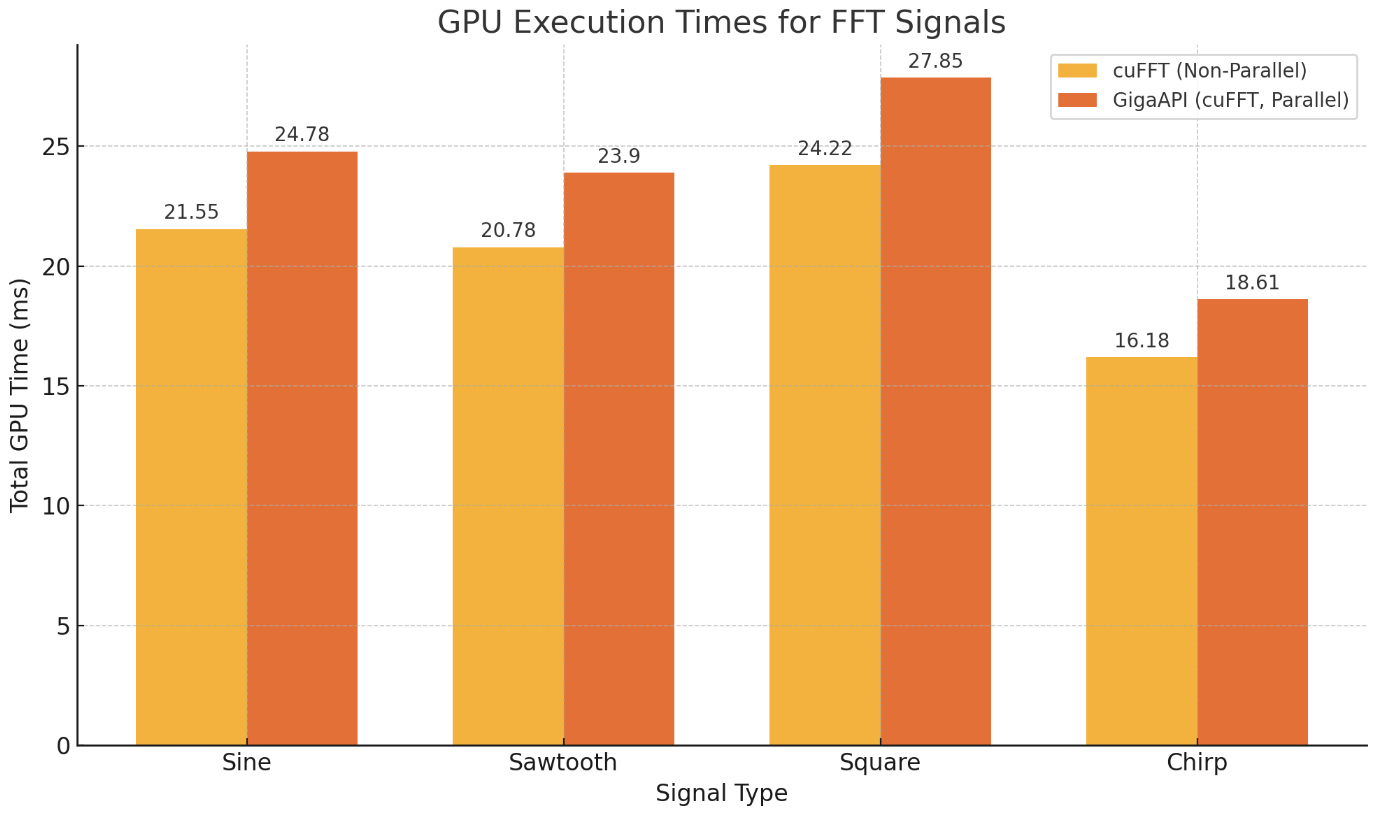}
    \caption{FFT Benchmark of Signals (cuFFT vs. GigaAPI)}
    \label{fig::fftbench}
\end{figure}

The performance of cuFFT on a single GPU surpassed that of our multi-GPU cuFFT implementation, mirroring the trend observed with cuBLAS. The relative closeness of our results can be attributed to the use of the optimized library, which partially offset the overhead costs associated with memory copies and other parallelism-related factors when processing signals with a frequency of 1.0 Hz, a sample rate of 1024.0 Hz, and a duration of 1.0 second. However, the single-GPU cuFFT ultimately proved superior, as it avoided the additional complexities inherent in multi-GPU setups while handling these specific signal parameters.

\subsection{API Benchmark \#2: Matrix Multiplication}
For matrix multiplication, we were very interested in how GigaAPI performs comparatively to cuBLAS. For reference, cuBLAS is a highly optimized CUDA API that is an implementation of BLAS (Basic Linear Algebra Subprograms) on top of the NVIDIA CUDA runtime. cuBLAS is very optimized and we ran through the CUDA source manual to get an understanding of how all the various functions work, especially how they interface with bare CUDA calls. A good thing to note here as a difference from GigaAPI and cuBLAS is that GigaAPI does not do as much error checking and thread safety as cuBLAS does: this leads to a bit of an overhead, especially on the smaller matrix-to-matrix multiplications with cuBLAS.

For our benchmark, we compare a bare cuBLAS implementation from scratch (using the user manual) against our GigaAPI implementation of matrix multiplication. We test sizes from $2^1$ all the way up to $2^{15}$, in which we generate random numbers and multiply matrices $A$ and $B$ together. We use CUDA events to record and relay the GPU usage stats to us, which we gather into a log file and then plot the results. It should be important to note here that we used a CUDA random number generator for both the cuBLAS and GigaAPI implementations: all numbers generated used the same random seed for reproducability.

Here are our results:
\begin{figure}[!ht]
    \centering
    \includegraphics[width=0.8\linewidth]{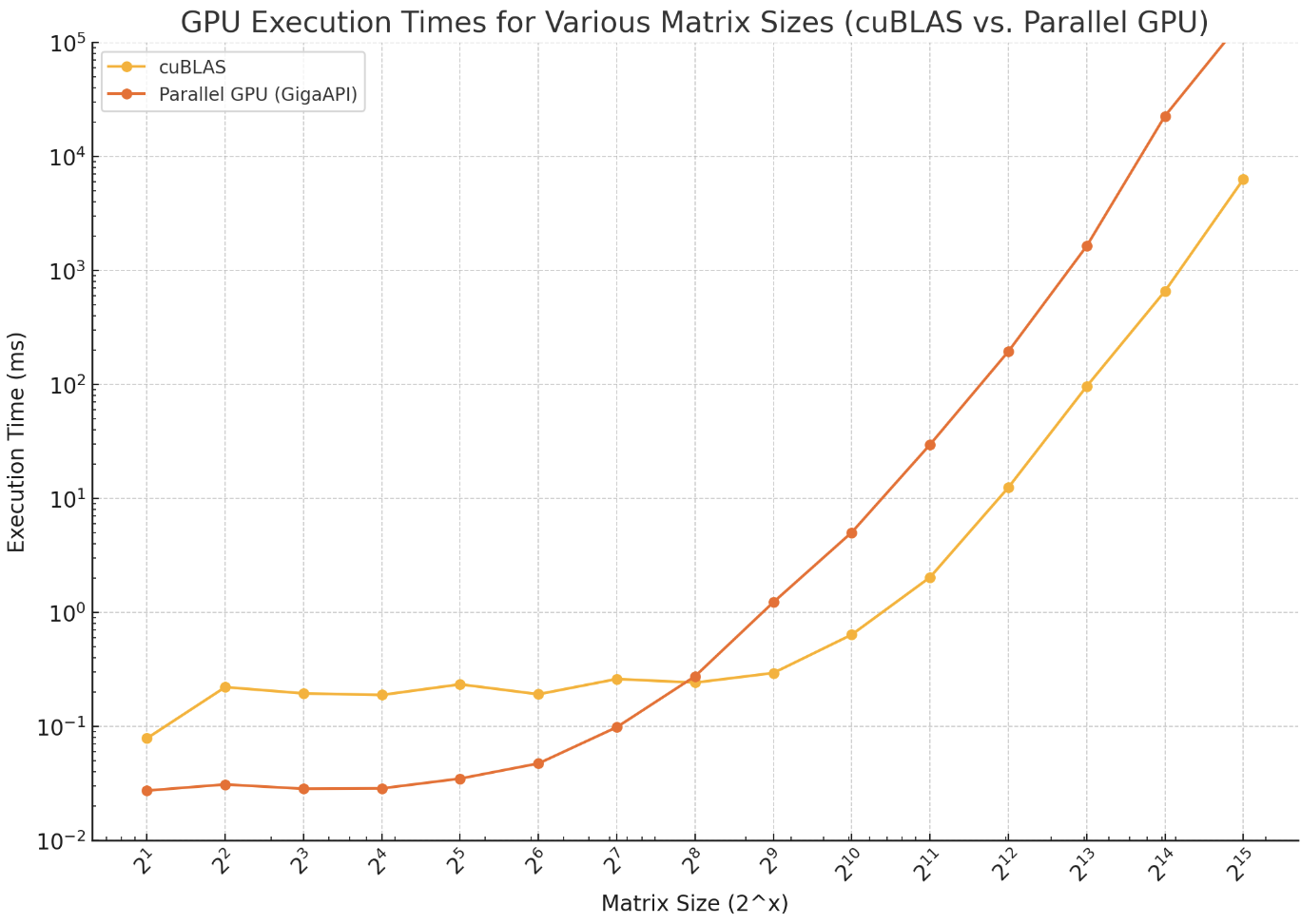}
    \caption{Matrix Multiplication Comparison (cuBLAS vs. GigaAPI)}
    \label{fig::matrixmult}
\end{figure}

We see some interesting results here. For random numbers filled inside matrices up to size $2^8$, GigaAPI actually outperforms cuBLAS (by a small factor). Again, as mentioned previously, this is most likely due to the large amounts of erroneous checks that cuBLAS implements, including checking every section of CUDA memory that is allocated. As we branch past this size, we begin to see a strong deviation, however, from GigaAPI and cuBLAS. At the very end of the peak, GigaAPI took 159 seconds to perform the matrix multiplication (despite the GPU parallelism), while cuBLAS only took about 6 seconds. This value is so incredibly large that it's actually cut off from the top of our graph: putting this point in would skew the results and not highlight the more minute differences for lower matrix sizes. Just to see what the results look like as the matrix size explodes for a better visualization, here is the same graph but with a linear scale instead:

\begin{figure}[!ht]
    \centering
    \includegraphics[width=0.8\linewidth]{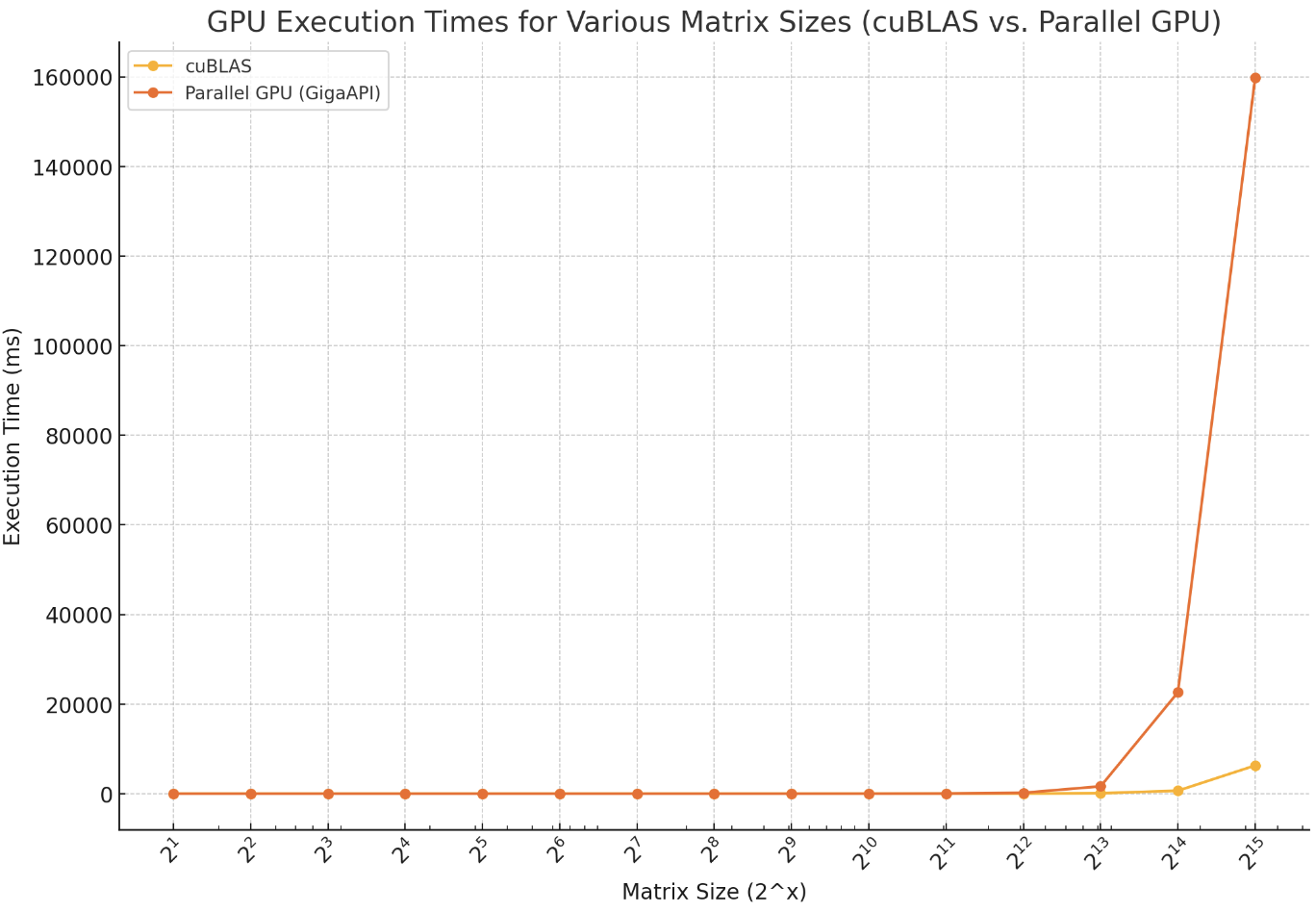}
    \caption{Linear representation of Matrix Multiplication (cuBLAS vs. GigaAPI)}
    \label{fig::linear_matrix_mult}
\end{figure}

So, it can be deduced that cuBLAS is definitely \textit{much} more optimized for matrices of larger sizes. After testing square matrices of 32768, we noticed that doubling the matrix size to $2^{16}$ was a ridiculous testing time: it was upwards of 10-15 mins per each run. This is a great situation of where despite existing parallelism, cuBLAS is already optimized to handle such situations. 

Digging deeper, we even saw that cuBLAS is maintained by a very extensive team of NVIDIA researchers, one being Roman Dubtsov. Dubtsov is the principal engineer for CUDA math-based libraries and has had about 15 years in the industry working on matrix multiplication within high-performance computing: our deviation started to make a little more sense after we found out about this.
\subsection{API Benchmark \#3: Vector Operations}

This was an interesting benchmark to run, as cuBLAS offers highly optimized and parallelized API functions. As aforementioned, our vector operations were solely limited to computing L2 norms and dot products, as we were interested in the effects that this would have on parallelization within GigaAPI. Our efforts were to split the computation evenly on both GPU's: while this is not the best use case for parallel GPU's at small vector sizes, it scales with complexity and becomes much more useful for extremely large vector sizes. 

As for our testbed, we decided to benchmark native cuBLAS vector L2 norm and dot product operations against GigaAPI. We expected that our results would be abysmal here: while our parallelized code was long and required context switching between both GPU's, the cuBLAS code for both the L2 norm and dot product was conducted in two total lines. The function headers were \verb|cublasSdot| and \verb|cublasSnrm2|: that is how easy it is to call cuBLAS functions on vectors. To really test how effective our code was, we chose to focus on extremely large vectors: we went all the way from $2^1$ vector size up to $2^{27}$ (approximated, actual vector size 67108864). Vectors were randomly filled with data on a standard distribution from -10 to 10, and here are our combined results for cuBLAS vs GigaAPI:

\begin{figure}[!ht]
    \centering
    \includegraphics[width=0.8\linewidth]{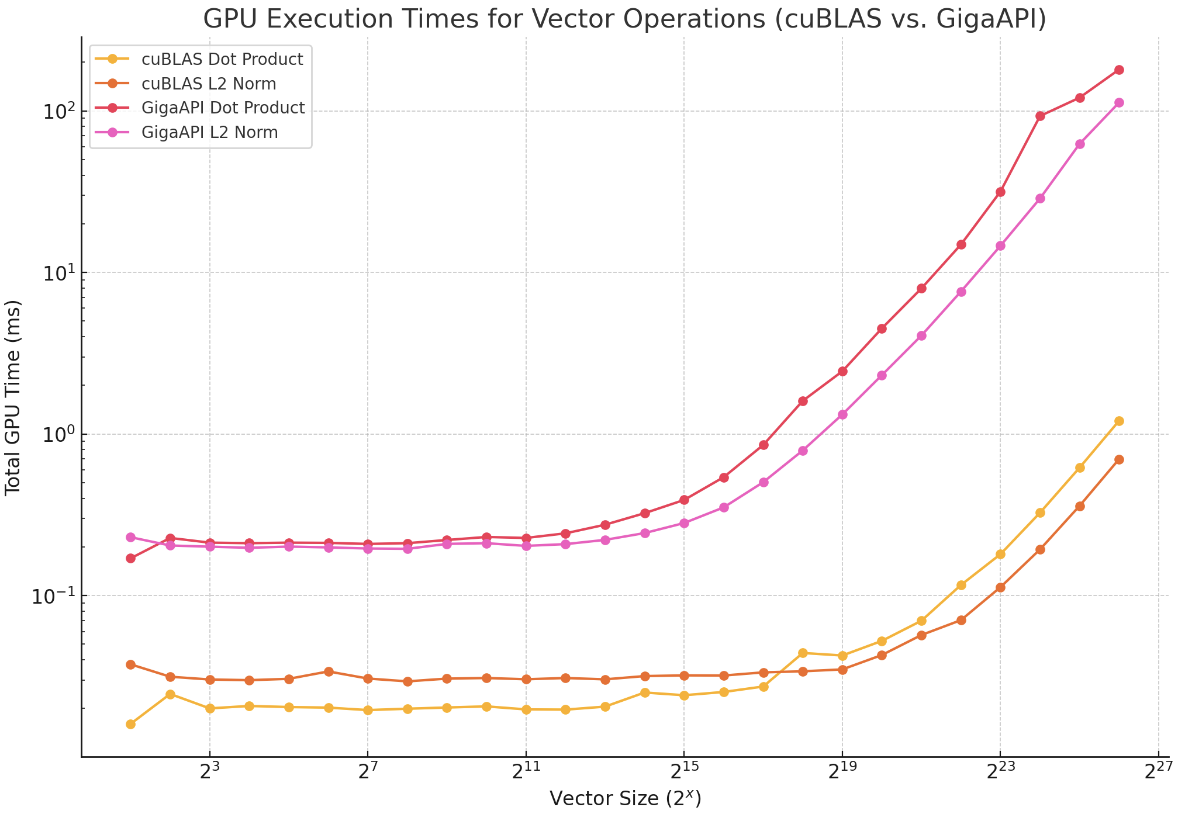}
    \caption{Vector Operations Benchmark (cuBLAS vs. GigaAPI)}
    \label{fig::vectorops}
\end{figure}

As you can see, GigaAPI is a good few orders of magnitude above the cuBLAS implementation, even despite our code being optimized for parallel GPU's. What's also interesting to see is that in both cuBLAS and GigaAPI the dot products took longer than the L2 norm: this could just be a way that internal CUDA libraries handle multiplication and addition over large rows. The delta between cuBLAS and GigaAPI shows just how powerful cuBLAS is as compared to our code: we even dug deeper into the source code, noting how highly optimized and efficient it was. 
\subsection{API Benchmark \#4: Upsampling}

Our first image suite benchmark involved testing how upsampling an image worked well in parallel, comparing upsampling a 4K image through scale factors of 2-40 (exiting when the GPU runs out of memory). We tested as follows: first loaded in a 4K image using OpenCV, then testing the single GPU load times for upsampling, and then testing our GigaAPI GPU load times for upsampling. 

One thing to note here: we spent a lot of time looking for a baseline upsampling kernel that was either provided by NVIDIA, or maybe even something open-source online used by industry. However, we struggled with finding useful code that worked without modification on our GigaGPU framework. So, to combat this, we ended up writing our own single GPU upsampling function: this just launches a single upsampling CUDA kernel on one GPU as opposed to having parallelization. It merely loads the full image into memory using CUDA, then runs the upsampling as per the blocks defined (same 16x16 structure).

We used the CUDA Runtime API to record and measure our events: we wrapped our kernel launches with CUDA recording events and printed out runtimes of GPU runtime per execution. One interesting thing to note here is that we had to have multiple CUDA recording events, as we had asynchronous memory copy, streams, and other stuff happening during our function: thus, we were very careful when recording events to ensure that we get nothing but the parallel GPU kernel execution times.

For our testbench, we upsampled a 4K base image from a factor of 2 all the way to 40. Here are our results:

\begin{figure}[!ht]
    \centering
    \includegraphics[width=0.8\linewidth]{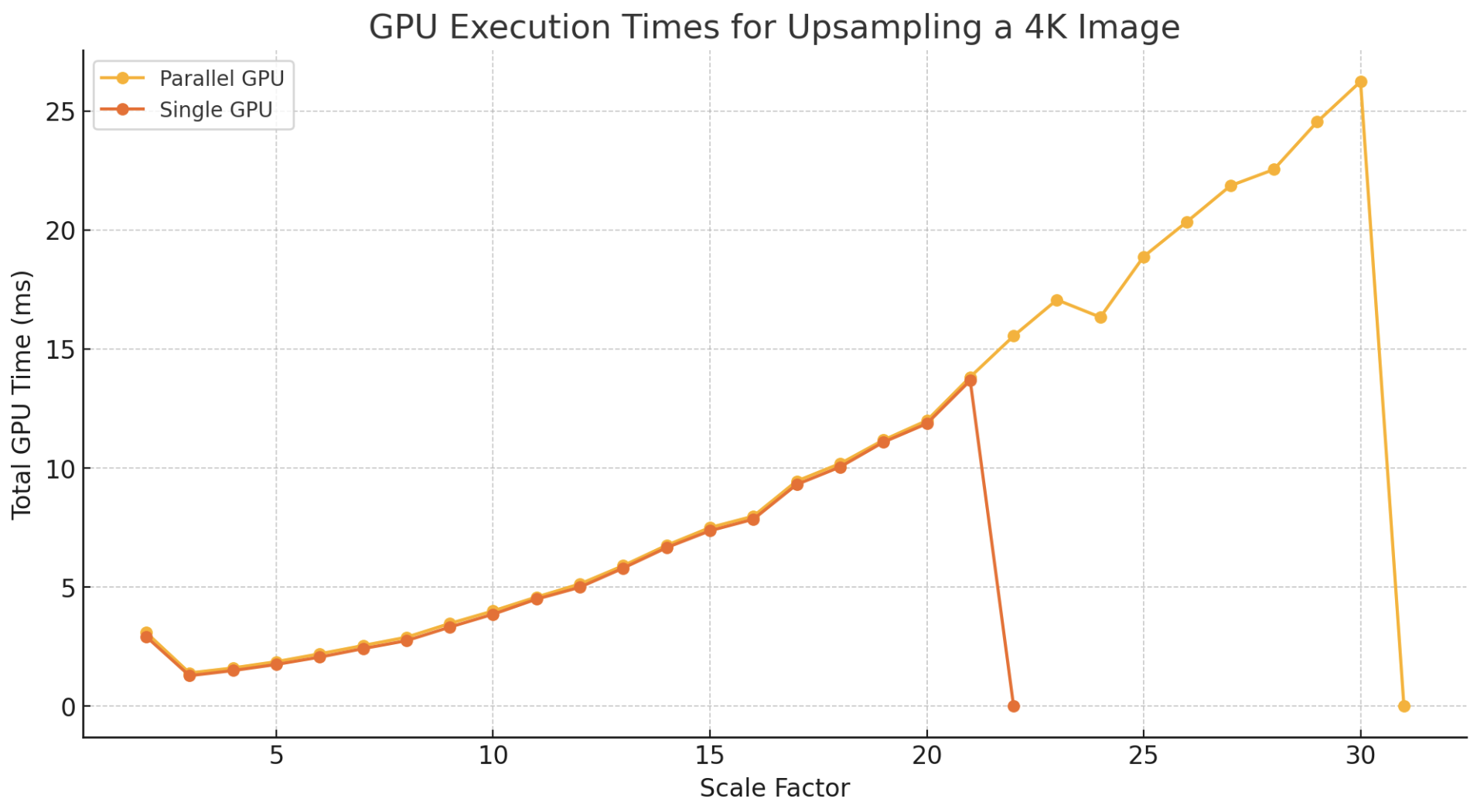}
    \caption{Upsampling Benchmark (Bare Kernel vs. GigaAPI)}
    \label{fig::upsampling}
\end{figure}

If you notice in the graph, there is a very interesting trend happening at 23 and at 32. For the single GPU, the upsampling actually segfaulted past a scale of 23 as the single GPU ran out of available memory to upscale the full image. However, on the parallel GPU using GigaAPI, we are able to go for a whole 9 more upsampling factors as the GPU data is batched and the pixels are split across both GPU's. Having the ability to compute in parallel keeps the times relatively straight as well, linearly increasing as opposed to exploding in complexity past some certain scale factor.

This is very good to see as it proves the need to have this parallel GPU setup for upsampling very large images. Suppose that we have a single (or double GPU, but not parallel) setup that is running upsampling on a large database. Assuming that execution of upsampling images is linear, if we hit a certain image quota, a segfault will kill the queue of remaining images, leaving them to not be processed and receiving errors on the user side. However, with a parallel GPU setup, having the GPU batch the pixels half and half to be upsampled is a good use and leaves much less of a chance of running into such errors, until the memory on both GPU's is exceeded.

\begin{figure}[!ht]
    \centering
    \includegraphics[width=0.8\linewidth]{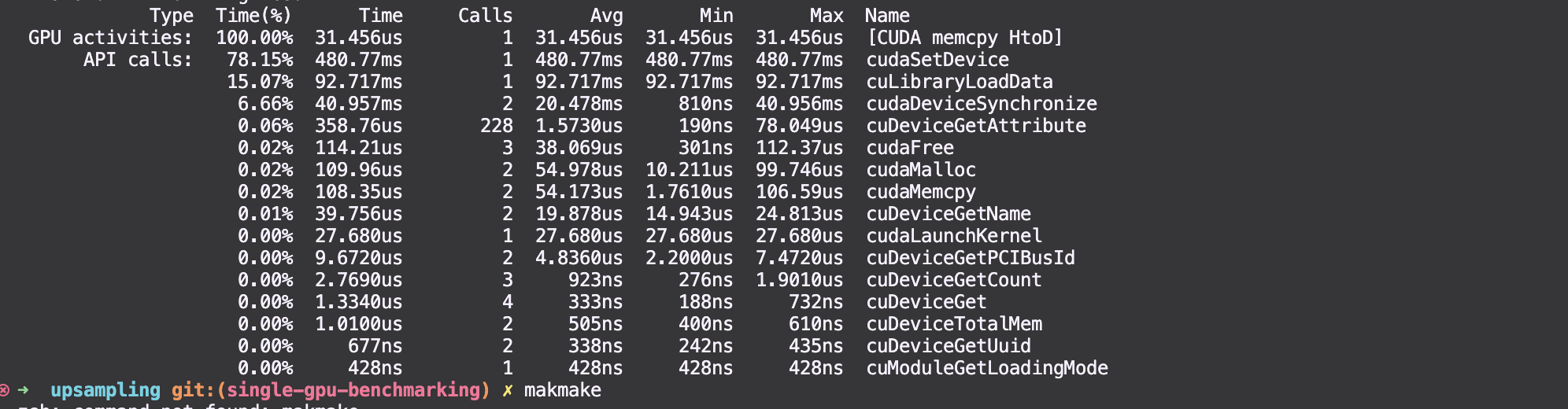}
    \caption{Single GPU \texttt{nvprof} Results}
    \label{fig::single_gpu}
\end{figure}

\begin{figure}[!ht]
    \centering
    \includegraphics[width=0.8\linewidth]{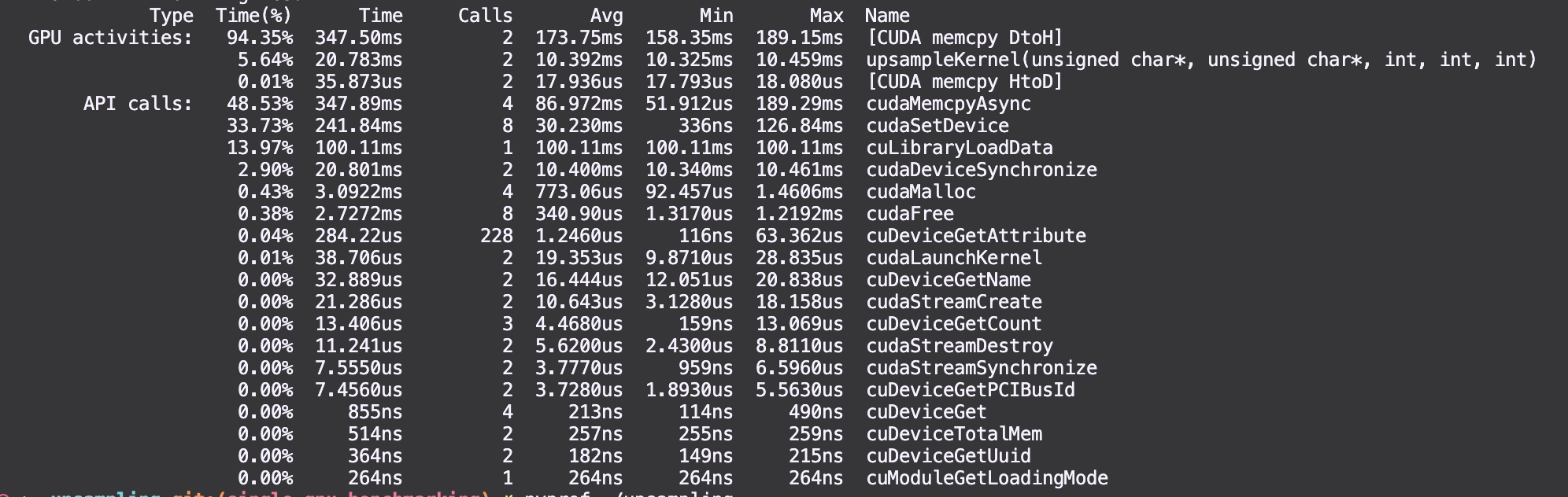}
    \caption{GigaAPI Parallel GPU \texttt{nvprof} Results}
    \label{fig::multi_gpu}
\end{figure}

We also used \verb|nvprof| to verify our results of upsampling. Again, we wanted to make sure that the kernels were launching in parallel for our GigaAPI and that only one kernel was launching for the single GPU, and we notice that in the debug output. See \ref{fig::multi_gpu} and \ref{fig::single_gpu} to compare these results and for a more in-depth look at all the kernel calls, API calls, and GPU utilization for this specific test.

\subsection{API Benchmark \#5: Sharpening}

In our benchmarking analysis of image sharpening, we implemented a function that performs image upsampling followed by sharpening on the upscaled image. The function was tested with scale factors up to 20. The results of our benchmarking led us to a clear realization: image sharpening using a Laplacian filter does not effectively showcase the advantages of parallelism when compared to more computationally intensive tasks.

\begin{figure}[!ht]
    \centering
    \includegraphics[width=0.8\linewidth]{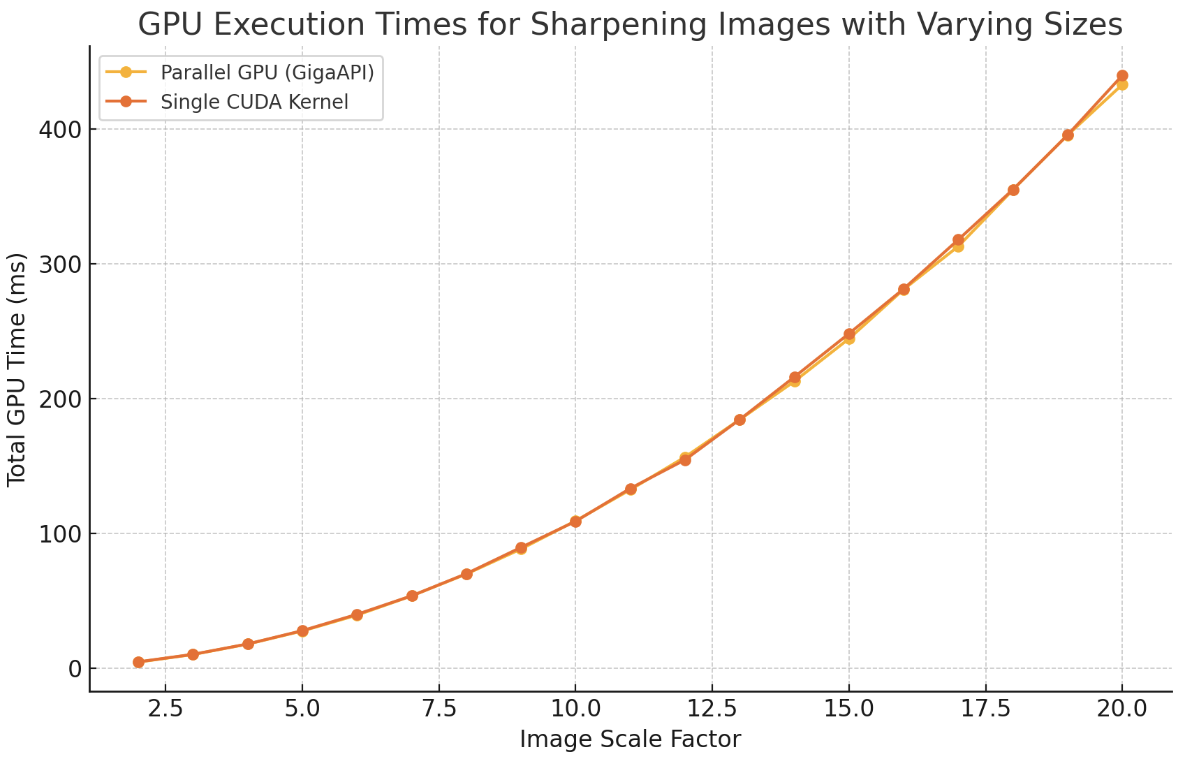}
    \caption{Sharpening an Image of Various Scales (4K)}
    \label{fig::sharpen}
\end{figure}

The Laplacian filter, being a simple 3x3 kernel, has a relatively low computational complexity when applied to an image. This inherent simplicity limits the potential benefits of parallelization. We hypothesized that for larger images, parallelism might yield some speedup in the sharpening process. However, even with an upsampling factor of 20, the performance remained largely unchanged between parallel and sequential implementations.

Based on our observations, we postulate that the advantages of parallelism in image sharpening would be more pronounced when employing more complex sharpening techniques. These techniques may involve computationally intensive operations, larger filter kernels, or additional processing steps. In such cases, the increased computational complexity would likely result in a more noticeable performance improvement when leveraging parallelism.

\subsection{API Benchmark \#6: RGB Grayscaling}

We also applied this benchmarking methodology to grayscaling RGB images, which involves converting a color image to a monochromatic representation. Similar to the Laplacian filter, grayscaling is computationally straightforward, and the performance difference between parallel and sequential implementations was minimal.

\section{Conclusions}
In conclusion, our extensive benchmarking and evaluation of GigaAPI, a user-space API designed for multi-GPU programming, has provided valuable insights into the potential and challenges of parallel GPU computing. Through our experiments, we have demonstrated the successful achievement of parallelism, as evidenced by the concurrent processing across multiple GPUs observed through NVTOP. 

Our benchmarking results have revealed that the performance of GigaAPI varies depending on the specific functionality being tested. In the case of FFT and matrix multiplication, GigaAPI showed competitive performance compared to cuFFT and cuBLAS for smaller sizes. However, as the matrix size increased, cuBLAS significantly outperformed GigaAPI, likely due to the highly optimized nature of cuBLAS and its maintenance by a dedicated team of NVIDIA researchers.

Similarly, for vector operations, cuBLAS demonstrated superior performance compared to GigaAPI, even for extremely large vector sizes. This highlights the efficiency and optimization of cuBLAS for vector computations and serves as a valuable lesson for us in terms of the level of optimization required to achieve high performance in parallel GPU computing.

Looking into image operations, our upsampling benchmark showcased the benefits of GigaAPI's parallel GPU setup. While the single GPU implementation segfaulted past a certain scale factor due to memory limitations, GigaAPI was able to handle larger upsampling factors by distributing the workload across both GPUs. GigaAPI has potential for memory-intensive tasks and this highlighted the importance of leveraging multiple GPUs to overcome memory constraints quite well.

For computationally simple tasks, our benchmarking of image sharpening and RGB grayscaling revealed that the advantages of parallelism may not be as pronounced. The performance difference between parallel and sequential implementations was minimal in these cases, indicating that the overhead of parallelization may outweigh the benefits for such tasks.

Throughout the development of GigaAPI, we encountered various limitations and issues, such as inconsistent error reporting and handling, high variability with CUDA code, and even lots of \verb|sudo| issues (as everything was in user-space). These limitations highlight the challenges in developing a robust and general-purpose user-space API for multi-GPU programming and serve as valuable lessons for future work in this area.

Overall, we feel that GigaAPI has provided us a foundation for understanding the potential and challenges of parallel GPU computing. While it demonstrates the feasibility of a user-space API for multi-GPU programming, further optimizations and improvements are necessary to achieve performance comparable to highly optimized libraries like cuBLAS/cuFFT/OpenCV. Our work also emphasizes the importance of considering the computational complexity of tasks when evaluating the benefits of parallelism.

As researchers, we acknowledge the limitations of our current implementation and the need for further research and development to address these challenges. We hope that our work will inspire others to build upon our findings and contribute to the advancement of parallel GPU computing, leading to more user-friendly APIs for multi-GPU programming!
\section{Source Code}

All of our source for this project can be found at: \\ 
\texttt{https://github.com/msuv08/parallel-gpus}. 

\section{Acknowledgements}

Special thanks to Professor Rossbach for teaching us the systems knowledge necessary to complete this project and for helping us scope our initial GPU programming idea into something more feasible and reasonable within the given time frame. Collectively, we spent north of 125 hours on this lab and are very grateful we had the time to delve deep and understand CUDA and user-space API's: these skills are quite useful to have and ones that we will sure to utilize in the future.

\bibliographystyle{plain}
\bibliography{references}

\end{document}